\documentclass[twocolumn]{revtex4}

\usepackage{amssymb}
\usepackage{graphics}
\begin{document}
\title{Entanglement Entropy in Extended Quantum Systems
\footnote{Plenary talk delivered at STATPHYS 23, Genoa, July
2007}}

\author{John Cardy}
\affiliation{Rudolph Peierls Centre for Theoretical Physics, 1
Keble Road, Oxford OX1 3NP, United Kingdom and All Souls College,
Oxford}
\date{August 2007}

\begin{abstract}
After a brief introduction to the concept of entanglement in
quantum systems, I apply these ideas to many-body systems and show
that the von Neumann entropy is an effective way of characterising
the entanglement between the degrees of freedom in different
regions of space. Close to a quantum phase transition it has
universal features which serve as a diagnostic of such phenomena.
In the second part I consider the unitary time evolution of such
systems following a `quantum quench' in which a parameter in the
hamiltonian is suddenly changed, and argue that finite regions
should effectively thermalise at late times, after interesting
transient effects.
\end{abstract}

\maketitle

\section{Introduction}
\label{intro} Entanglement is one of the most fundamental features
of quantum mechanics, yet in some ways the most mysterious. In
this talk I will discuss some of the recent developments in
understanding its role in many-body systems extended in space,
focussing on two main topics: the universal properties of
entanglement entropy near quantum critical points, and the
behaviour of entanglement, and, more generally, correlation
functions, after what is termed a `quantum quench'.

Most of the work I will be presenting was carried out in
collaboration with P.~Calabrese \cite{cc1,cc2,cc3} and others. For
a more general perspective the reader is referred to a recent
review by Amico \em et al. \em\cite{amico}.
\section{Entanglement entropy near quantum critical points}
\subsection{Bipartite quantum entanglement}
\label{sec:1} 
\begin{figure}
\centering
\resizebox{0.65\columnwidth}{!}{%
  \includegraphics{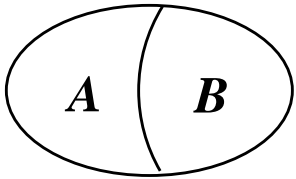}}
\caption{Schematic depiction of the decomposition of the Hilbert
space. Later these will also represent regions of $d$-dimensional
space.}
\label{fig1}       
\end{figure}

Consider a general quantum system prepared in a pure
state $|\Psi\rangle$, so that it has density matrix
$\rho=|\Psi\rangle\langle\Psi|$. We suppose that the Hilbert space
can be written as a direct product ${\cal H}={\cal
H}_A\otimes{\cal H}_B$ (see Fig.~\ref{fig1}).
We imagine two observers (traditionally named Alice and Bob) such
that Alice can make observations only in ${\cal H}_A$ (that is
corresponding to linear operators of the form ${\cal O}_A\otimes
1_B$), and correspondingly for Bob. In general Alice's
observations are entangled with those of Bob.

One of the most useful mathematical results in understanding how
to quantify entanglement is that of Schmidt decomposition, which
is based on the property of singular value decomposition for
matrices. It states that any pure state $|\Psi\rangle$ may be
written as
$$
|\Psi\rangle=\sum_jc_j|\psi_j\rangle_A\otimes|\psi_j\rangle_B\,,
$$
where $|\psi_j\rangle_{A,B}$ are orthonormal states in ${\cal
H}_A$ and ${\cal H}_B$ respectively, and $\sum_j|c_j|^2=1$.
Moreover the $c_j$ can be chosen to be real and $\geq0$. (Note
that there is only one sum here: for each state in ${\cal H}_A$
there is just one state in ${\cal H}_B$.)

One measure of the entanglement in $|\Psi\rangle$ is the \em
entropy \em
$$
S\equiv-\sum_jc_j^2\log c_j^2\,.
$$
If $c_1=1$ and all the rest vanish, $|\Psi\rangle$ is a product
state and is unentangled (although there may still be classical
correlations.) If, on the other hand, all the $c_j$ are equal,
then $S$ takes its maximal value, given by the logarithm of the
smaller of the dimensions of ${\cal H}_A$ and ${\cal H}_B$. For
example, if each space is a direct product of $N$ qubits (spin
$\frac12$ degrees of freedom) then the maximal entanglement
entropy is $N\log2$.

Equivalently, we can define the entanglement entropy as the von
Neumann entropy $S_A=-\mbox{Tr}_A\rho_A\log\rho_A$ of Alice's
density matrix $\rho_A=\mbox{Tr}_B\rho$. Evidently $S_A=S_B=S$.

Other measures of entanglement exist \cite{amico}, but the entropy
has several nice properties: additivity, convexity, basis
independence. In quantum information theory \cite{bennett}, it
gives the maximum efficiency of conversion of partially entangled
to maximally entangled states that Alice can achieve using only
local operations in her part of the Hilbert space. From a
computational point of view, it gives the amount of classical
information required to specify the reduced density matrix
$\rho_A$. This important, for example, in the density matrix
renormalisation group (DMRG).

However, in this talk I want to use the entanglement entropy as a
basis-independent means of characterising quantum phase
transitions. This is particularly important in those cases where
it is not clear what the order parameter is, or which correlation
functions should become long-ranged at the transition. (Indeed,
for topological transitions there may be no identifiable order
parameter.) I shall consider the case when the degrees of freedom
of the quantum system are distributed over some large region $\cal
R$ in $d$-dimensional euclidean space, and the hamiltonian $H$
contains only short-range interactions, for example a quantum spin
system, or, more generally a UV-cutoff quantum field theory. (I
shall focus solely on the universal properties near the phase
transition: these, as K.~Wilson and others taught us, are all
encoded in the field theory description.) The subspace ${\cal
H}_A$ will consist of the degrees of freedom in some large
(compact) subset A of $\cal R$, and we shall assume that the whole
system is in a pure state, usually the ground state $|0\rangle$ of
$H$. However, it will be useful also to consider the case when the
whole system is in a thermal mixed state with $\rho\propto
e^{-\beta H}$.

The main question I will address is how the entanglement entropy
depends on the size and geometry of the region A, and on the
universality class of the critical behaviour.

\subsection{Entanglement entropy from the path integral}
\begin{figure}
\centering
\resizebox{0.90\columnwidth}{!}{%
  \includegraphics{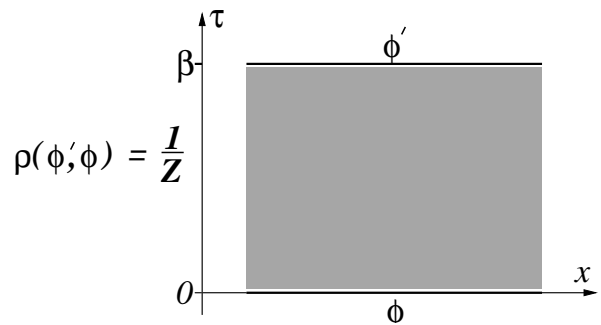}}
\caption{The path integral in space $x$ and imaginary time $\tau$.
The rows and columns of the density matrix are labelled by the
values of the fields (degrees of freedom) at $\tau=0$ and $\beta$
respectively.}
\label{fig2}       
\end{figure}

\begin{figure}
\centering
\resizebox{0.6\columnwidth}{!}{%
  \includegraphics{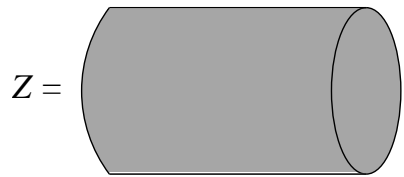}}
\caption{The partition function $Z$ is obtained by sewing together
the top and bottom edges of the world sheet in Fig.~\ref{fig2},
that is by identifying the fields and integrating them.}
\label{fig3}       
\end{figure}

\begin{figure}
\centering
\resizebox{0.75\columnwidth}{!}{%
  \includegraphics{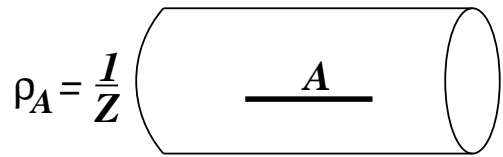}}
\caption{The reduced density matrix is obtained by sewing together
the top and bottom edges of the world sheet in Fig.~\ref{fig2},
along only the parts corresponding to region B.}
\label{fig4}       
\end{figure}

\begin{figure}
\centering
\resizebox{0.5\columnwidth}{!}{%
  \includegraphics{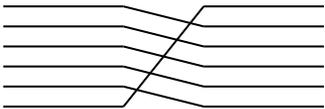}}
\caption{A view along the edges A of $n$ copies of
Fig.~\ref{fig4}, showing how they are to be sewn together to give
$\mbox{Tr}\rho_A^n$.}
\label{fig5}       
\end{figure}

\begin{center}
\em ``Quantum mechanics is just statistical mechanics in one more
dimension'' \em (M.E.~Fisher)
\end{center}
From an analytic perspective, the Schmidt decomposition is very
difficult to carry out except in simple cases, and we employ a
different route via the path integral. As Feynman taught us, the
thermal density matrix $\propto e^{-\beta H}$ may be written as a
path integral in an imaginary time interval $(0,\beta)$
(throughout we adopt units with $\hbar=1$.) This is illustrated in
Fig.~\ref{fig2}.
The density matrix is correctly normalised by the partition
function $Z$ as shown in Fig.~\ref{fig3}.
However, we need the reduced density matrix $\rho_A$. This is
found by sewing together only those parts of the upper and lower
edges corresponding to the region B (Fig.~\ref{fig4}).
In order to compute the entropy $S_A=-\mbox{Tr}_A\rho_A\log\rho_A$
we use a device reminiscent of the `replica trick' in disordered
systems: we compute $\mbox{Tr}\rho_A^n$ for positive integral $n$,
analytically continue in $n$, and evaluate
$$
S_A=-\mbox{Tr}\,\rho_A\log\rho_A=-\left.\frac\partial{\partial
n}\right|_{n=1}\mbox{Tr}\,\rho_A^n\,.
$$
For positive integer $n$, this is given by taking $n$ copies of
Fig.~\ref{fig4} and sewing the edges along A in a cyclic fashion,
as shown in Fig.~\ref{fig5}.
For $d=1$ (which is the case we shall consider here) this
expresses $\mbox{Tr}\rho_A^n$ as a path integral on an $n$-sheeted
Riemann surface, with branch points at the ends of the interval A
(if A has several disjoint parts, there are several branch
points.)
\subsubsection{High temperature limit}
\begin{figure}
\centering
\resizebox{0.75\columnwidth}{!}{%
  \includegraphics{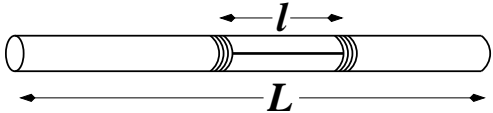}}
\caption{The world sheet for the path integral for
$\mbox{Tr}\rho_A^n$ in the high temperature limit. In the central
section, corresponding to the region A, it winds around $n$
times.}
\label{fig6}       
\end{figure}

Before proceeding further, it worth understanding how this
formalism gives the expected answer at high temperatures. Consider
the case where A is an interval of length $\ell$ in a system of
total length $L$, with $\ell\ll L$. Further suppose that
$\beta\ll\ell$. The path integral is then over a very narrow
cylinder (compared to its length) as shown in Fig.~\ref{fig6}.
In this limit we expect the result of the path integral to
factorise:
$$
Z_n\approx Z_1(\ell,n\beta)Z_1(L-\ell,\beta)^n\,,
$$
where $Z_1(L,\beta)$ is the partition function for a system of
length $L$ at inverse temperature $\beta$. Thus
$$
\mbox{Tr}\rho_A^n\sim\frac{Z_1(\ell,\beta)}{Z_1(\ell,\beta)^n}\sim
\frac{\exp{-n\beta F_A(n\beta)}}{\exp(-n\beta F_A(\beta))}
$$
where $F_A(\beta)$ is the usual Helmholtz free energy of region A.
Differentiating with respect to $n$ at $n=1$, we then find
$$
S_A\sim\beta(E_A-F_A)\,.
$$
This shows that in this limit, the von Neumann entropy becomes the
usual thermodynamic entropy, as expected.
\subsection{The critical case in $d=1$}
Suppose the 1$d$ system is at a quantum critical point with
dynamic exponent $z=1$ (that is, linear dispersion relation
$\omega=v|k|$ at low energies). Then dimensional analysis (with
$\hbar=v=1$) implies that $F_A(\beta)\sim -\pi c\ell/6\beta^2$.
This is just the one-dimensional version of Stefan's law: the
constant $c=1$ for a single species of boson. We conclude that for
$\ell\gg\beta$
$$
\mbox{Tr}\rho_A^n\sim\exp\left[-\frac{\pi
c}{6\beta}\left(n-\frac1n\right)\ell\right]\,.
$$
The above considerations do not appear to shed much light on the
case we wish to consider, namely zero temperature. However, in
this case the limit $\ell\gg\beta$ is related to the opposite one
$\beta\to\infty$ by \em conformal symmetry\em: the mapping
$z\to(\beta/2\pi)\log z$ converts exponential decay along the
cylinder into power law decay at $T=0$. Thus, at zero temperature
we have
$$
\mbox{Tr}\rho_A^n\sim \ell^{-(c/6)(n-1/n)}\,.
$$
Taking the derivative we see that
\begin{equation}
\label{SA} S_A\sim(c/3)\log\ell\,.
\end{equation}
For a general conformal field theory (CFT), $c$ is the central
charge. (\ref{SA}) was first found in 1994 by Holzhey \em et al.
\em\cite{holzhey}, who called it the geometric entropy.

Note that this implies that, even at a quantum critical point, the
entropy grows only logarithmically in the length $\ell$, as
opposed to its maximum allowed behaviour which is $O(\ell)$. This
accounts for the success of the DMRG method in $d=1$. This
logarithmic growth is apparently not restricted to critical points
with $z=1$: for example it also holds in random spin chains
\cite{refael}.

(\ref{SA}) is just one example of a plethora of universal results
which can be found using CFT methods. For example, one can explore
the cross-over between finite and zero temperature when
$\ell\sim\beta$ to find the elegant formula \cite{korepin,cc1}
$$
S_A\sim(c/3)\log\big((\beta/\pi)\sinh(\pi\ell/\beta)\big)\,.
$$

\subsection{Finite correlation length in $d=1$}
\begin{figure}
\centering
\resizebox{0.75\columnwidth}{!}{%
  \includegraphics{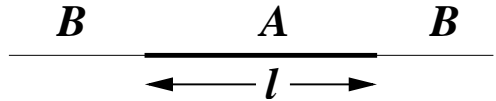}}
\caption{An interval A of length $\ell$. When the correlation
length $\xi\ll\ell$, the entanglement arises from regions of
length $O(\xi)$ near the boundaries between A and B.}
\label{fig7}       
\end{figure}

The above CFT considerations apply only at the quantum critical
point. However, the behaviour of the entanglement entropy close to
such a point may be deduced from this by scaling arguments. For
example, consider the case when A is an interval of length $\ell$
in an infinite system (Fig.~\ref{fig7}).
When the correlation length $\xi$ is finite and $\ll\ell$, we
expect that the entanglement arises from regions of length
$O(\xi)$ near the boundaries between A and B. Since the scaling at
the critical point is logarithmic we may therefore conjecture
that, in this limit,
\begin{equation}
\label{SA2} S_A\sim 2\times(c/6)\log\xi\,,
\end{equation}
and that in general there is a contribution $(c/6)\log\xi$ from
each contact point between A and B. For the case of a single such
point, when A and B are both semi-infinite lines, this has been
derived exactly using the corner transfer matrix
\cite{cc1,peschel}, for a large class of integrable models.
Recently the leading corrections to (\ref{SA2}) have been computed
\cite{CCD}. These are of the form $O(e^{-2\ell/\xi})$ and they
appear to be rather universal.

\subsection{Higher dimensions $d>1$}
In higher dimensions, the natural conjecture for the entanglement
entropy (given that $S_A$ should equal $S_B$) is that it is
proportional to the `area' $|\partial A|$ of the boundary between
the two regions. This has been verified in various models
(although logarithmic factors are also possible) and even proved
rigorously \cite{arealaw} starting from reasonable assumptions
about the behaviour of correlations. On dimensional grounds the
coefficient of the area law should go as $a^{1-d}$ where is $a$ is
the short-distance cutoff, and therefore be non-universal.
However, renormalisation group arguments similar to those used to
analyse the scaling behaviour of the free energy imply that there
should be a universal term proportional to $\xi^{1-d}|\partial A|$
hidden behind this \cite{cc1}.

It is interesting to note that these area-dependent terms may
cancel in more complicated entanglement combinations. For example,
if A and B are different subregions of $\cal R$, then one may
conjecture that
$$
S_{A\cup B}+S_{A\cap B}-S_A-S_B
$$
is universal, and, at the critical point, depends only on the
geometry and some universal constants of the critical theory.

\section{Time-dependence after a quantum quench}
In the second part of this talk I want to discuss a subject which
is intimately related to entanglement properties of quantum
many-body systems, but is somewhat more general. Let us suppose
that we prepare a system at time $t=0$ in a pure state
$|\Psi_0\rangle$, which we usually take to be the ground state of
some translationally invariant hamiltonian $H_0$ with a gap $m_0$
to the first excited state. For times $t>0$ we then evolve the
state according to a \em different \em hamiltonian $H$ (which
doesn't commute with $H_0$). Note that this evolution is \em
unitary\em: no dissipation or noise. This protocol has been termed
a `quantum quench'.

We may then ask how the reduced density matrix $\rho_A$ of some
finite part of the system, its entropy $S_A$ and correlation
functions of local operators ${\cal O}(x)$ with $x\in A$ evolve.
In particular, does $\rho_A$ reach a stationary state, and if so
how is it characterised?

For traditional solid state systems the assumption of unitary
evolution usually breaks down so rapidly that these kind of
questions have hardly been addressed in the past, except in a few
cases of integrable spin chains \cite{mccoy}. However recent
experiments on cold atoms in optical lattices have shown that it
is possible to maintain coherence over measurable time intervals.
This has prompted more recent work, both in integrable systems and
more generally \cite{sengupta,cc2,cc3,peschel2}. These detailed
considerations all lead to the prediction of the following rather
simple physical effects.

\subsection{The horizon effect}
\begin{figure}
\centering
\resizebox{0.9\columnwidth}{!}{%
  \includegraphics{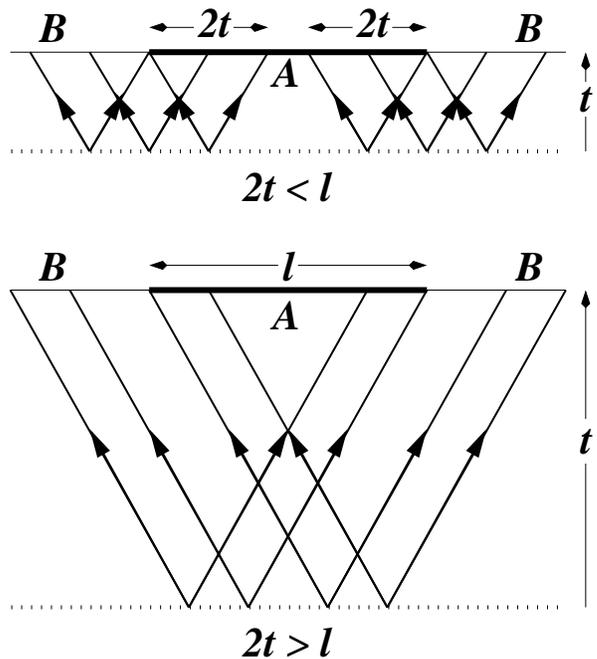}}
\caption{Quasiparticles are emitted at $t=0$, and subsequently
move classically. Left and right-moving pairs of particles are
entangled, and if they arrive on A and B respectively they cause
entanglement of these two regions.}
\label{fig8}       
\end{figure}

Since $|\Psi_0\rangle$ has (extensively) higher energy than the
ground state of $H$, it acts as a source of quasiparticles of $H$.
The analysis of simple models shows that subsequently they move
classically. This is illustrated in Fig.~\ref{fig8}, for the
simplest case in $d=1$ when all particles travel at the same speed
$v$.
Entanglement between regions A and B (which is very small
initially) arises when two entangled particles emitted from nearby
points arrive on A and B respectively. From the upper diagram in
Fig.~\ref{fig8}, corresponding to early times $t<\ell/2v$, we see
that the number of such pairs of particles, and therefore the
degree of entanglement, increases linearly with $t$. One the other
hand for $t>\ell/2v$ is saturates at a value proportional to
$\ell$. This is what is found in explicit calculations. The
coefficient of the linear term depends on the initial state.
\subsubsection{Correlation functions}
\begin{figure}
\centering
\resizebox{0.6\columnwidth}{!}{%
  \includegraphics{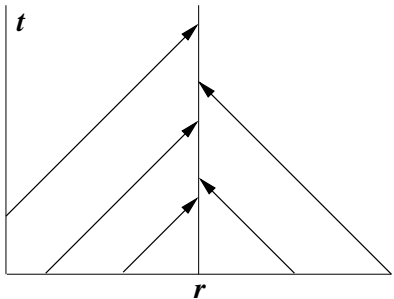}}
\caption{A local observable, for example a quantum spin, is
subject to incoherent radiation of quasiparticles emitted from
different spatial points. Its value therefore flips according to a
Poisson process, and its expectation value relaxes exponentially.}
\label{fig9}       
\end{figure}

In general it is found that one-point functions $\langle{\cal
O}(x,t)\rangle$ of local operators (for example spins in a quantum
spin chain where $|\Psi_0\rangle$ breaks the spin-reversal
symmetry but $H$ and its ground state do not), decay exponentially
fast towards their values in the ground state of $H$. The physical
reason for this is illustrated in Fig.~\ref{fig9}.
On the other hand the behaviour of a two-point function
$\langle{\cal O}(x_1,t){\cal O}(x_2,t)\rangle$ is governed by the
horizon effect. Up to time $t\sim |x_2-x_1|/2v$ its connected part
does not change from its initial form which is very short-ranged.
For wider separations $\langle{\cal O}(x_1,t){\cal
O}(x_2,t)\rangle\sim\langle{\cal O}(t)\rangle^2$, which decays
exponentially in time. At the time $t\sim|x_2-x_1|/2v$ the two
points fall inside the horizon and the connected correlation
function becomes non-zero. In the case when all quasiparticles
have the same speed, the full correlation function then becomes
time-independent, and, since it was previously decaying
exponentially with $t$, now exhibits exponential decay in the
separation $|x_2-x_1|$.
\subsubsection{General dispersion relation}
\begin{figure}
\centering
\resizebox{0.75\columnwidth}{!}{%
  \includegraphics{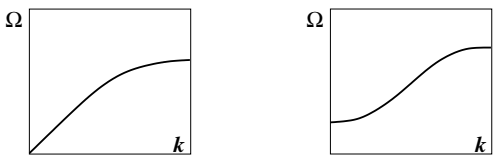}}
\caption{Two examples of dispersion relations. In each case, the
quasiparticles move with their group velocity given by the slope
of this curve. The fastest moving particles are responsible for
the horizon effect, and the slowest particles for the approach to
the limiting behaviour at late times.}
\label{fig10}       
\end{figure}

These physical considerations allow us to understand the behaviour
in more realistic situations where lattice effects or a gap modify
the dispersion relation. Examples are shown in Fig.~\ref{fig10}.
The fact that the quasiparticles, once emitted, travel classically
means that it is their group velocity which is important. Since
this can vanish, for example at the zone boundary, this is
responsible for a very slow (power law + oscillations) approach to
the asymptotic limit at large $t$, which is actually seen in exact
calculations and numerical work.
\subsection{Thermalisation at late times}
We have argued that calculations in various solvable models
suggest that correlation functions for $x\in A$, and by
implication the reduced density matrix $\rho_A$, become stationary
at late times. But what is their form? In these cases it turns out
that, at least in the case where the gap $m_0$ in the spectrum of
$H_0$ is large, $\rho_A$ has a \em thermal \em form $\propto
e^{-\beta_{\rm eff}H}$, where $\beta_{\rm eff}^{-1}$ is an
effective temperature dependent on $m_0$ among other things.

In order to understand this, first consider a very simple example:
a simple harmonic oscillator quenched from frequency $\omega_0$
(and initially in the ground state), to frequency $\omega$. It is
a simple calculation to evaluate the overlap between the ground
state $|\Psi_0\rangle$ and a typical eigenstate of $H$ with energy
$E$. For $\omega_0\gg\omega$ we find
$$
\langle\Psi_0|E\rangle\propto\exp(-\beta_{\rm eff}E/2),,
$$
where $\beta_{\rm eff}\sim 4/\omega_0$, so the matrix elements of
the time-dependent density matrix have the form
$$
\langle E|\rho(t)|E'\rangle\sim e^{-\beta_{\rm eff}(E+E')/2} \,
e^{i(E-E')t}\,.
$$
Note that since the energy differences are all multiples of
$\omega$, $\rho(t)$ does not tend towards a stationary value -- it
oscillates as expected!

However in an extensive system, which we can consider as a set of
oscillators, one for each quasiparticle mode $k$, if we consider a
finite region A of size $\ll t/v$ as $t\to\infty$, we need to
integrate over all the $k$-modes. If the dispersion relation has
the form $\omega_k=m+O(k^2)$ the different modes destructively
interfere except at $k=0$. The result is that correlation
functions, and $\rho_A(t)$, become stationary as $t\to\infty$ as
if they were at finite temperature. In fact a more careful
calculation, valid not just in the large $m_0$ limit, gives
$$
\beta_{\rm eff}=(4/m)\tanh^{-1}(m/m_0)\,.
$$
Similarly, we find that the extensive part of $S_A$ saturates at a
value equal to the thermodynamic entropy at this effective
temperature.

Although the above simple argument relied on a non-zero gap $m>0$,
in fact (as long as interactions are present) the above result
appears to hold (for example in a CFT in 1+1 dimensions) even in
the gapless case.

It should be stressed that this effective thermalisation occurs
despite the fact that the system as a whole remains in a pure
state, and there is no ergodicity or coupling to a heat bath --
the effect arises solely as a consequence of quantum interference
and entanglement.

\subsection*{Summary}
In this talk I have argued (a) that entanglement entropy provides
a useful order-parameter independent diagnostic of quantum phase
transitions, with many universal features, and (b) that after a
quantum quench, there are interesting transient phenomena like the
horizon effect, and that at late times finite regions should
behave as though they are in thermal equilibrium.

There are many open questions, particularly in regard to part (b),
and whether it holds only for theories which admit a quasiparticle
picture. In addition, these kinds of calculations need to be
repeated for more realistic models, and also for quenches through
the critical point into an ordered phase, where the relation to
later-time coarsening effects needs to be understood.

\noindent\em Acknowledgements. \em This work was supported in part
by EPSRC grants GR/R83712/01 and EP/D050952/1. I thank Pasquale
Calabrese for numerous discussions.


\begin{thebibliography}{}

\bibitem{cc1} P.~Calabrese and J.~Cardy, J. Stat. Mech. \textbf{0406}, (2004) P06002.
\bibitem{cc2} P.~Calabrese and J.~Cardy, J. Stat. Mech. textbf{0504}, (2005) P04010.
\bibitem{cc3} P.~Calabrese and J.~Cardy, Phys. Rev. Lett. \textbf{96}, (2006)
136801; J. Stat. Mech. \textbf{0706}, (2007) P008.
\bibitem{amico} L.~Amico, R.~Fazio, A.~Osterloh and V.~Vedral,
arXiv:quant-ph/0703044.
\bibitem{bennett} C.H.~Bennett, H.J.~Bernstein, S.~Popescu and
B.~Schumacher, Phys. Rev. A \textbf{53}, (1996) 2046.
\bibitem{holzhey} C.~Holzhey, F.~Larsen and F.~Wilczek, Nucl.
Phys. B \textbf{424}, (1994) 44.
\bibitem{refael} G.~Refael and J.E.~Moore, Phys. Rev. Lett.
\textbf{93}, (2001) 260602.
\bibitem{korepin} V.~Korepin, Phys. Rev. Lett. \textbf{92}, (2004)
096402.
\bibitem{peschel} I.~Peschel, J. Stat. Mech. \textbf{0412}, (2004)
P12005.
\bibitem{CCD} J.~Cardy, O.~Castro-Alvaredo and B.~Doyon,
arXiv:0706.3384.
\bibitem{arealaw} M.M.~Wolf, F.~Verstraete, M.B.~Hastings and
J.I.~Cirac, arXiv:0704.3906.
\bibitem{mccoy} E.~Barouch and B.~McCoy, Phys. Rev. A \textbf{2},
(1970) 1075; \textbf{3}, (1971) 786; \textbf{3}, (1971) 2137.
\bibitem{sengupta} K.~Sengupta, S.~Powell and S.~Sachdev, Phys.
Rev. A \textbf{69}, (2004) 053616.
\bibitem{peschel2} V. Eisler and I. Peschel, J. Stat. Mech.
\textbf{0706}, (2007) P06005.


\end{thebibliography}
\end{document}